\documentstyle[aps,prl,twocolumn,epsfig]{revtex}

\begin{document}

%\wideabs{

\title{Anomalous rotational properties of Bose-Einstein condensates in asymmetric traps}

\author{Juan J. Garc\'{\i}a-Ripoll, V\'{\i}ctor M. P\'erez-Garc\'{\i}a}

\address{Departamento de Matem\'aticas, Escuela T\'ecnica Superior de
  Ingenieros Industriales, \\
  Universidad de Castilla-La Mancha 13071 Ciudad Real, Spain}

\date{\today}

\maketitle

\begin{abstract}
  We study the rotational properties of a Bose-Einstein condensate confined in
  a rotating harmonic trap for different trap anisotropies.  With simple
  arguments, we obtain the velocity field of the quantum fluid for condensates
  with or without vortices.  While the condensate describes open spiraling
  trajectories, on the frame of reference of the rotating trap the fluid moves
  against the trap's rotation. We also find expressions for the angular
  momentum and linear and Thomas-Fermi solutions for a vortex-less state. In
  these two limits we find the same analytic relation between the shape of the
  cloud and the rotation speed. Our predictions are supported by numerical
  simulations of the mean field Gross-Pitaevskii model.
\end{abstract}


%}

%03.75.Fi Phase coherent atomic ensemble (Bose condensation)
%05.30.Jp Boson systems
%67.57.De Superflow and hydrodynamics
%67.57.Fg Textures and vortices (superfluids)

\begin{figure}
  \label{fig-velocities}
  \center{
  \epsfig{file=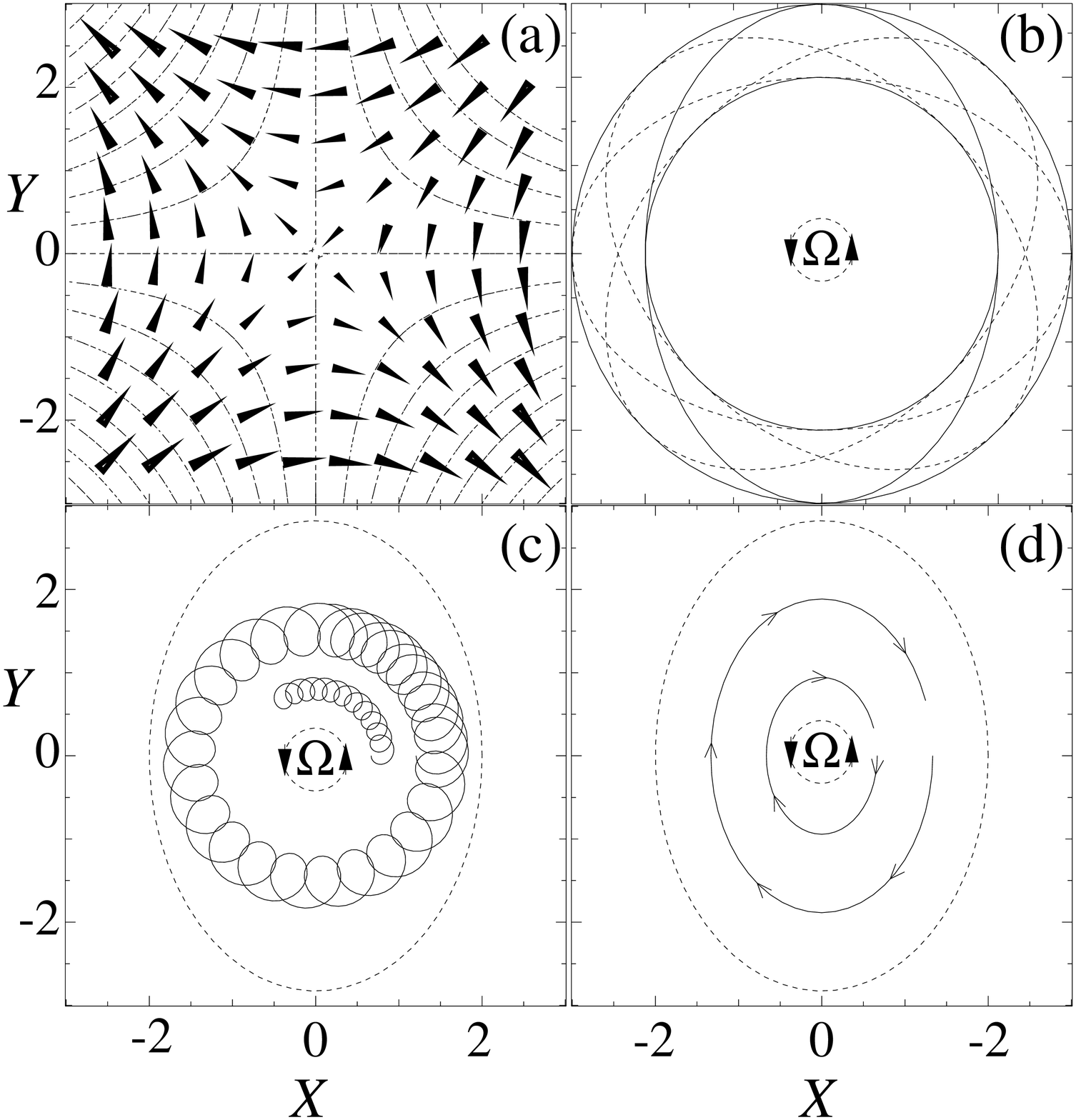,width=0.9\linewidth}}
  \caption{
    Flow features of an asymmetric vortex-less condensate with form factor
    $a\simeq2$ and rotation speed $\Omega=1$. (a) Velocity distribution at
    $t=0$ (b) The rotation of the asymmetric condensate defines two circles
    which define the width of the spiraling flow in the following picture. (c)
    Flow on the laboratory frame: the shape of the cloud (dashed line) in the
    Thomas-Fermi limit, and the paths of two test particles -- inner spirals
    for 6 rotation and outer spiral for 20 rotation periods (1 period is
    $2\pi/\Omega$). (d) Flow on the rotating frame using the same case as in
    (c) but for only one period. The motion of the particles is opposite to the
    trap rotation.}
\end{figure}

The question of whether an atomic gas may be a superfluid and the quest for
distinct superfluid properties are among the central goals of recent research
with atomic Bose-Einstein condensates (BEC). These targets have been largely
inspired by previous achievements on $^4$He and its superfluid phase He-II, so
that from one known property of ``classical" superfluids some analog is found
for the ``new" condensates. To name a few phenomena which have been found this
way we can cite the discovery of anomalous sound velocities
\cite{Zaremba,MIT-sound}, the discovery of a critical velocity for the
beginning of viscous damping \cite{MIT-damping}, the study of vortices
\cite{Rokhsar-Nature,Vortices-many,Fetter,Vortices-Feder,Vortices-nuestro} and their
generation \cite{Mattews,Vortices-rotan} and the study of rotational properties
such as scissors modes \cite{Scissors,Oxford-scissors} or moments of inertia
\cite{Stringari-inertia}.

In this context there has been a great interest in achieving Bose-Einstein
condensation with rotating traps, a task which has been completed recently
\cite{Vortices-rotan,Oxford-scissors}.  The motivation of this interest is that
the ground state of a condensate in a rotating trap can be forced to host one
or more vortices depending on the angular speed
\cite{Rokhsar-Nature,Vortices-nuestro}, thus providing us with persistent
currents which are themselves traces of superfluidity.  Although most of the
theoretical work regarding the generation and stability of vortices has focused
on isotropic traps \cite{Vortices-many,Rokhsar-Nature,Vortices-nuestro}, it is
only using truly anisotropic traps that one may expect the condensate to offer
a mechanical response to the rotation and produce a vortex. This idea is
developed throughout the paper.

There are few relevant results related to non-isotropic traps. First
\cite{Stringari-inertia}, the moment of inertia of an inhomogeneous condensate
have been studied assuming that the gas is an ideal one, i. e. {\em without
  interactions}. Among the most striking results is the formula which relates
the moment of inertia to the asymmetry of the condensate given by the
expectation value $\langle x^2-y^2\rangle$.  In Ref. \cite{Fetter} the
dynamics and stability of a vortex in this type of traps is studied. And finally in Refs.
\cite{Vortices-Feder} the ground state of a condensate in a rotating asymmetric
trap is found numerically for particular values of the rotation speed and their
stability is studied.

In this paper we study the rotational properties of a condensate in an rotating
anisotropic trap, with and without a vortex. We consider the problem within the
mean field approximation, thus taking interactions into account (an essential
difference with previous work \cite{Stringari-inertia}).  Our treatment is both
analytic and numeric: we obtain expressions for the most relevant observables,
which are exact in some limits, and verify them with complex numerical
simulations over a large range of parameters. The main predictions of the paper
are {\em the explicit expressions for the moments of inertia} and a
characterization of the flow of the quantum fluid, whose main features are
reflected in Fig. 1.

{\em The model.-}
We will study a two-dimensional condensate in a harmonic trap whose axes rotate
at a certain angular speed, $\Omega$. The potential may be described by
\begin{equation}
V\left({\bf x},t\right) =V_0\left( e^{-\Omega tR_{z}}{\bf x}\right) =V_0\left( U\left
    ( -\Omega t\right) {\bf x}\right),
\end{equation}
where $R_z$ is the generator of the rotations in the $\Re ^2$ space,
$U(\theta)$ is an orthogonal transformation which rotates a vector an angle
$\theta$ around the origin, and
$V_0(r_1,r_2)=\frac{1}{2}\left( \omega _1^2r_1^2+\omega _2^2r_2^2\right)$.

We will work on two different frames of reference: the laboratory frame
$\{S,{\bf x}\}$ which is stationary and the rotating frame
$\{\tilde{S}, {\bf r}=U(-\Omega t){\bf x}\}$ which moves with the
trap. On the first frame the zero temperature mean field theory reads
\begin{equation}
\label{GPE-full}
i\partial _{t}\psi \left( x,t\right) =
\left[ -\frac{1}{2}\triangle _{x}+V\left( x,t\right) + g\left| \psi \right| ^2\right] \psi \left( x,t\right).
\end{equation}
Here $g$ characterizes the interaction and is defined in terms of the ground
state scattering length. It is important to remark that our solutions have a
well defined norm that we may take as equal to the number of particles
\cite{nota}, $\Vert \psi \Vert^2=\Vert \phi \Vert^2=N$. To simplify the
treatment we have assumed a set of units in which $\hbar=m=1$.

We will define a second wave function, this time on $\tilde{S}$, given by
$\phi({\bf r},t) =\psi(U(\Omega t){\bf r},t)$ with ${\bf r} = (r_1, r_2)$. The
evolution of this function is ruled by
\begin{equation}
\label{GPE-rotating}
i\partial _{t}\phi =
\left[ -\frac{1}{2}\triangle _{r}+V_0\left( {\bf r}\right) + g\left| \phi \right| ^2-\Omega L_z\right] \phi.
\end{equation}
Here $L_z=-i\left( r_1\partial_2\phi -r_2\partial_1\phi \right)$ is a
representation of the angular momentum operator along the z-axis.

{\em Stationary states.-} Let us write Eq. (\ref{GPE-full}) in the
modulus-phase representation in the rotating frame of reference. Defining
($\phi = \sqrt{\rho} e^{i\Theta}$) we obtain the continuity equation for the
density
\begin{equation}
\label{continuity}
\partial_t \rho-\Omega \nabla \rho\cdot\left(R_z{\bf r}\right)=-\nabla \cdot \left[ \rho {\bf v}\right],
\end{equation}
where ${\bf v}=\hbox{Im} (\bar{\phi}\nabla \phi) =\nabla \arg
\phi =\nabla \Theta $ is almost the velocity field of the quantum fluid in the
stationary frame. Actually the velocity frame on $S$ is given by
${\bf V}=U\left(-\Omega t\right) {\bf v}\left( U\left( -\Omega t\right)
\right)$.

Eq. (\ref{GPE-rotating}) has a set of stationary states, which are solutions of
the type $\phi({\bf r},t) =e^{-i\mu t}\phi({\bf r},0).$ These states represent
configurations of the condensed cloud which maintain their shape and move
rigidly with the trap (keep in mind that ${\bf r}$ corresponds to the rotating
system).

In our paper we will be interested in the lowest energy state, the ground
state. \emph{The main assumption} throughout the paper will be that the lines
of constant density of a ground state are ellipses of an unknown shape, i. e.,
\begin{equation}
\label{hypothesis}
\rho =\rho\left( u^2\right) =\rho\left(r_1^2+r_2^2/a\right),
\end{equation}
which is exact for the ground state in certain limits to be discussed later.  We will
also need a normalized anisotropy factor, $k=(1-a)/(1+a)\in [-1,+1]$.

Using Eqs (\ref{hypothesis}) and (\ref{continuity}), one finds
\begin{equation}
h\left( u\right) \left[ \Omega \left( 1-a\right) r_1r_2
  + ar_1 \partial_1 + r_2 \partial_2\right] \Theta = \triangle \Theta ,
\end{equation}
where $h={\text d}\ln\rho /{\text d}(u^2).$ Let us look for solutions
corresponding to an incompressible flow, i.e. $\triangle \Theta = 0$. Then
\begin{equation}
\label{maclaro}
\Omega \left[ \left( 1-a\right) r_1r_2\right] + \left[ ar_1 \partial_1 + r_2 \partial_2\right]  \Theta   =  0,
\end{equation}
Writing the solution of Eq. (\ref{maclaro}) in the form
\begin{equation}
\label{Theta-general}
\Theta(r_1,r_2,t) = -\mu t+\Omega \frac{a-1}{a+1} r_1 r_2+ \Theta _{vort}(r_1,r_2),
\end{equation}
it is clear that there is still an undetermined part of the phase,
$\Theta_{vort}$ which satisfies $\triangle \Theta_{vort} = 0$ and $\nabla
\Theta_{vort} \perp \nabla \rho$. This phase varies along elliptic paths
surrounding the origin and carries vorticity in case there is any.

Once we have found the phase, the density may be obtained by solving
\begin{equation}
\label{GPE-density}
\mu =-\frac{\triangle \sqrt{\rho }}{2\sqrt{\rho }}+ \frac{\left( \nabla \Theta \right) ^2}{2} + V_0+g\rho + i \Omega L_z \Theta .
\end{equation}
This can be done analytically both in the non-interacting limit, $gN
\rightarrow 0$ and on the Thomas-Fermi limit, $gN \rightarrow \infty$ as we
will show later.

{\em States without vorticity.-}
When the BEC ground state has no vorticity we can make $\Theta_{vort} = 0$
in Eq. (\ref{Theta-general})
\begin{equation}
\label{Theta-nodeless}
\Theta =-\mu t+\Omega \frac{a-1}{a+1}r_1r_2.
\end{equation}
As it was already expected, in the radially symmetric case ($a=0$) the solution
is also radially symmetric, the phase is uniform and the velocity field becomes
zero.

For an asymmetric trap the velocity field is {\em different from zero but still
  irrotacional}, $\nabla \times {\bf V}=0.$ The definition of ${\bf V}$ above
provides us with analytic expressions for the flow on the laboratory. This flow
is a product of two rotations around the origin: a circular rotation with
angular speed $\Omega$ and an elliptic rotation with a smaller angular speed
$\omega=-\Omega\sqrt{1-k^2}$ on the opposite sense! This is a striking two-fold
result.  First, on the trap frame $\tilde{S}$ the fluid is moving along closed
elliptic contours opposite to the rotation of the trap [See Fig. 1(d)] but with
a smaller velocity. And second, the composition of both movements on the
laboratory gives a flow made of slow and typically open spiraling paths [See
Fig. 1(c)].

Roughly speaking, the fact that the actual flow on the laboratory frame is much
slower is a signature that the inertia of a condensate is smaller than the
inertia of a classical fluid. We will prove this calculating the angular
momentum of the cloud, which is given by a simple expression
\begin{equation}
\label{Lz-m0}
\left\langle L_z\right\rangle _{ground}
= \int \rho \left( r\times \nabla \Theta\right)_z
= \Omega \frac{\left\langle r_1^2-r_2^2\right\rangle ^2}{\left\langle r_1^2+r_2^2\right\rangle }.
\end{equation}
Here we have used $a=\left\langle r_1^2\right\rangle /\left\langle
  r_2^2\right\rangle.$ As it was expected, the moment of inertia,
$I=\left\langle L_z\right\rangle/\Omega$, becomes zero for a symmetric trap.
Furthermore, it is always smaller than the classical value $I=\langle
r_1^2+r_2^2 \rangle$. We must also remark that that Eq. (\ref{Lz-m0}) differs
from the zero temperature limit of Ref. \cite{Stringari-inertia} because our
method is not perturbative and allows the shape of the cloud to depend on the
rotation speed.

Inserting Eq. (\ref{Theta-nodeless}) into Eq. (\ref{GPE-rotating}) we find
the equation of a nonlinear harmonic oscillator with a pair of effective
frequencies which depend on $\Omega$,
\begin{mathletters}
\label{TF-freqs}
\begin{eqnarray}
\omega_{x,eff}^2 & = & \omega _1^2+\Omega ^2 k (k-2),\\
\omega_{y,eff}^2 & = & \omega _2^2+\Omega ^2 k (k+2),
\end{eqnarray}
\end{mathletters}
We can recover the asymmetry of the cloud from the exact solutions of this
oscillator on two different limits, the non-interacting or {\em linear} limit,
$g \rightarrow 0$, and the Thomas-Fermi limit. Both expressions are
\begin{mathletters}
\label{TF-a}
\begin{eqnarray}
a_{L} &=& a\left( g \rightarrow 0 \right) = \omega_{x,eff}/\omega_{y,eff}, \\
a_{TF} &=& a\left( g \rightarrow \infty \right) = \left(\omega_{x,eff}/\omega_{y,eff}\right)^2.
\end{eqnarray}
\end{mathletters}
It is apparent in Eq. (\ref{TF-a}) that rotation emphasizes the anisotropy. In
Fig. 2(c-d) we compare the Thomas-Fermi and linear approximations with two
different numerical experiments with an anisotropic trap. The main conclusion
is that Eq.  (\ref{TF-a}(b)) works extremely well for medium to large $N$.
There is a small error which depends on the norm of the solution. This error is
related to the fact that a Thomas-Fermi approximation wipes out any dependence
of the shape of the cloud on the the number of particles, $N$.

{\em Nucleation of the first vortex.-} As the rotation speed is increased the
fluid adapts by evolving to different states, which may involve the nucleation
of one or more vortices. These vortices arise both as zeros of the density and
as additive contributions, $\Theta_{vortex}$, to the phase of the cloud. The
phase of the vortices suffers a discontinuous change around any closed
circuit which encloses them. Hence vortices satisfy Feynmann's condition for
the quantization of the superfluid flow
\begin{equation}
\oint \nabla \Theta _{vortex} \cdot \text{d}{\bf l}=2\pi m,\quad m=0,\pm 1,\pm 2\ldots 
\end{equation}
Following the reasoning developed throughout this paper, we should now solve a
Poisson equation for $\Theta_{vortex}$ with some restrictions on the direction
of $\nabla \Theta_{vortex}$. Instead we will follow a more intuitive path which
gives accurate results. In the framework of our approximation $\Theta_{vortex}$
will correspond to a flow around the lines of constant density of the cloud, as
it happens in the symmetric case. In elliptic polar coordinates, $\{r_1 = u\cos
\theta, r_2 = \sqrt{a}u\sin \theta\}$ the gradient of the phase becomes
\begin{equation}
\nabla \Theta _{vortex} \simeq g(u)\left( -\sin \theta ,\sqrt{a}\cos \theta \right),
\end{equation}
where $g(u)$ is a decreasing function of the radius of each ellipse. This
expression may be integrated along elliptic contours to find the dependence
of the vortex phase on the elliptic angle
\begin{equation}
\label{Theta-vortex}
\Theta_{vortex}(\theta)\simeq m\theta +m\frac{a-1}{2\left( a+1\right) }\sin 2\theta.
\end{equation}
We have found the typical dependence for a symmetric vortex, $m\theta$, plus a
correction which comes from the asymmetry. Using Eq. (\ref{Theta-vortex}) we
estimate the new value of the angular momentum. It has two contributions, one
coming from the rotation of the cloud and another one coming from the vortex,
\begin{equation}
\label{Lz-m1}
\left\langle L_z\right\rangle =
\Omega \frac{\left\langle r_1^2-r_2^2\right\rangle ^2}{\left\langle
    r_1^2+r_2^2\right\rangle }+
\frac{2m\sqrt{\langle r_1^2 \rangle \langle r_2^2 \rangle}}{\langle r_1^2+r_2^2 \rangle}N.
\end{equation}
In the symmetric trap case the introduction of a vortex implies a fixed change
of the total angular momentum $\Delta L_z=\left\langle L_z\right\rangle
_{vortex}-\left\langle L_z\right\rangle _{ground}=N$. The asymmetry radically
changes this picture and the more asymmetric the trap is the smaller the amount
angular momentum kept in the vortex -- a quantity which eventually becomes zero
in the $|k|\rightarrow 1$ limit.

\begin{figure}
  \label{fig-predict}
  \center{\epsfig{file=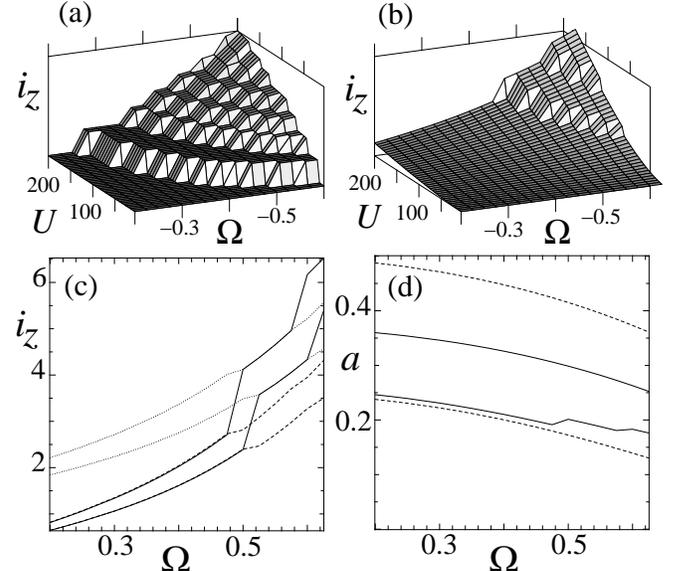,width=1\linewidth}}
  \caption{
    (a-b) Evolution of the moment of inertia per particle $i_z=L_z/N\Omega$ as
    a function of the rotation speed and $N$ for (a) a radially symmetric trap
    and (b) an asymmetric trap with $\omega_2^2=2\omega_1^2$. (c) Moment of
    inertia versus the angular speed of the asymmetric trap (solid line),
    estimation for a vortex-less state (dotted line) and estimation for a state
    with a single vortex (dashed line). (d) Asymmetry of the cloud, $a$, in the
    asymmetric trap as theoretical prediction (dotted line), numerical result
    for $N=5$ (upper solid line) and $N=175$ (lower solid line). All figures
    are in adimensional units.}
\end{figure}

{\em Numerical simulations.-} Up to this point we have obtained several
predictions concerning the stationary states of a BEC in a rotating trap.
Although there are no experimental results on this yet, we can contrast our
theoretical work with numerical simulations of the Gross-Pitaevskii equation (GPE).

First let us recall that the stationary solutions of Eq. (\ref{GPE-rotating})
are critical points of the functional \cite{Vortices-nuestro}
\begin{equation}
\label{functional}
E[\phi] =\int \bar{\phi }\left( -\frac{1}{2}\triangle _{r}+V_0\left( r\right) +\frac{g}{2}\left| \phi \right| ^2-\Omega L_z\right) \phi,
\end{equation}
subject to the constraint of a fixed norm $\int \left| \phi \right|
^2d^{n}r=N$.  Some of the critical points of Eq. (\ref{functional}) correspond
to ground states. These states represent the typical configuration of the gas
when it is Bose-condensed in the rotating trap. To find the ground states we
have minimized Eq. (\ref{functional}) using the Sobolev's gradients method in a
discrete Fourier basis. This was performed for a two-dimensional radially
symmetric trap, $\omega _1=\omega _2$, and for an asymmetric one, $\omega
_1=\sqrt{2}\omega _2$, varying the only relevant parameters, $\Omega$ and $N$,
over a wide range. As a result we obtained two maps which show the angular
momentum of the ground state as a function of $(\Omega,N)$, and
which are plotted in Fig. 2(a-b).

In Fig. 2(a) we recover the discontinuous distribution which was found in Ref.
\cite{Rokhsar-Nature}. Here the system starts with zero angular momentum and
remains still until a certain angular speed, $\Omega_1(N)$. Beyond
$\Omega_1(N)$ a vortex grows and we have another plateau on which the angular
momentum remains constant, $L_z=N$. Once more there is a critical frequency,
$\Omega_2(N)$, which marks the nucleation of another vortex. From then on the
evolution of $L_z$ is a piecewise differentiable one.  There are still more
jumps due to the creation of more vortices, but $L_z$ is no longer constant in
each interval, since the vortices may move to accommodate more angular
momentum.

In Fig. 2(b) the picture is quite different. First the angular
momentum has a non-trivial dependence on the number of particles and on the
angular speed. It is neither zero for a vortex-less state, nor constant for a
state with one vortex. Indeed, the dependence with respect to $\Omega$ is
extremely well reproduced by Eq. (\ref{Lz-m0}) and by Eq. (\ref{Lz-m1}), which
suggests that these are exact laws which could be derived by some other mean.

{\em Conclusions.-} We have achieved several goals in this work. First, we have
shown how to obtain analytic results about the macroscopic properties of a
rotating condensate, which leads to unexpected and intuitively appealing
results for the flow of the condensate.  And second, we have derived precise
laws which relate the anisotropy of the condensate, its moment of inertia and
the two relevant parameters of the problem: the norm, $N$, and the trap angular
speed, $\Omega$. Our predictions have been numerically confirmed generating
maps of $L_z$, $\langle x^2 \rangle$ and $\langle y^2 \rangle$ as a function of
$(\Omega,N)$.

We hope that these methods will support further work, such as the study of
linear stability of vortices in asymmetric condensates, the development of
better approximations to the shape of the cloud or even a variational analysis
which describes the transition between different vorticities. Furthermore, our
predictions are easily extensible to three-dimensional condensates giving many
results (condensate shape, moment of inertia or energy release, for instance)
which could be tested in current experiments \cite{Oxford-scissors}.  For
instance, the discontinuity on the asymmetry of the cloud when a vortex is
nucleated could be used as a {\em non-destructive} mean to ensure the existence
of a phase singularity in the atomic cloud.

This work has been partially supported by CICYT under grant PB96-0534.


\begin{references}
\bibitem{nota}{For a three-dimensional setup with an almost two-dimensional
    condensate, $N$ is not exactly the number of particles but rather the limit
    of a three-dimensional one. Several criteria may be used to connect $N$ in
    the 2D case to the real number of particles.}
\end{references}
\end{document}